\input{aipcheck}
\documentclass[
    ,final            % use final for the camera ready runs
  ]
  {aipproc}
\layoutstyle{6x9}
\usepackage{subfigure}

\catcode`\|=11
\let\= = \equiv
\def\_#1{_{\!#1}}
\def\||#1{{\left\Vert#1\right\Vert^{\kern-0.5em\phantom{1}}_{\kern-0.5em\phantom{1}}}}
\def\|#1{{\left\vert#1\right\vert^{\kern-0.5em\phantom{1}}_{\kern-0.5em\phantom{1}}}}
\def\(#1{{\left(#1\right)^{\kern-0.5em\phantom{1}}_{\kern-0.5em\phantom{1}}}}
\def\[#1{{\left[#1\right]^{\kern-0.5em\phantom{1}}_{\kern-0.5em\phantom{1}}}}
\def\{#1{{\left\lbrace#1\rbrace^{\kern-0.5em\phantom{1}}_{\kern-0.5em\phantom{1}}}}

\def\cases#1{\def\\{\cr}
          \left\lbrace\>\vcenter{\normalbaselines\m@th
 \ialign{$##\hfil$&&\quad$##\hfil$\crcr#1\crcr}}\right.^{\kern-0.5em\phantom{1}}_{\kern-0.5em\phantom{1}}}

\def\frac#1#2{{#1\over#2}}

\mathchardef\gamma="710D %§ Questa deve precedere tutte le macro che utilizzano la c.s. \gamma
\mathchardef\Gamma="7100 %§ Questa deve precedere tutte le macro che utilizzano la c.s. \Gamma

\mathchardef\lambda="7115 %ß
\mathchardef\Lambda="7103 %ß

\def\rbrace{\right\}}

\def\sqr#1#2{{\vcenter{\vbox{\hrule height.#2pt
 \hbox{\vrule width.#2pt height#1pt \kern#1pt
 \vrule width.#2pt}
 \hrule height.#2pt}}}
}

\def\Tag#1{\ifmmode\eqno{(#1)}\hbox to\rightskip{\null}\else(#1)\fi}

\begin{document}

\title{Measuring cosmological distances by coalescing binaries}
\classification{04.25.dg, 04.25.dk, 04.30.-w, 98.80.Es}
\keywords{Gravitational waves, standard candles, cosmological distances.}

\author{I. De Martino}{address={F\'{\i}sica Te\'orica, Universidad de Salamanca, 37008 Salamanca, Spain}}
\author{S. Capozziello}{address={Dipartimento di Scienze Fisiche, Universit\`a di Napoli "Federico II" and
INFN sez. di Napoli Compl. Univ. di Monte S. Angelo, Edificio G, Via Cinthia, I-80126 - Napoli, Italy}}
\author{M. De Laurentis}{address={Dipartimento di Scienze Fisiche, Universit\`a di Napoli "Federico II" and
INFN sez. di Napoli Compl. Univ. di Monte S. Angelo, Edificio G, Via Cinthia, I-80126 - Napoli, Italy}}
\author{M. Formisano}{address={Dipartimento di Fisica, Universit\`a di Roma "La Sapienza", Piazzale Aldo Moro 5, I-00185 Roma, Italy and 
INAF-IAPS, Via Fosso del Cavaliere 100, 00133 Roma (Italy) }}

\date{\today}

\begin{abstract}
Gravitational waves detected  from well-localized inspiraling
binaries would allow us to determine, directly and independently,
binary  luminosity  and  redshift. In this case, such
systems could behave as "standard candles"  providing  an
excellent probe of cosmic distances up to $z <0.1$ and complementing other indicators of cosmological distance ladder.
\end{abstract}
\maketitle

%%%%%%%%%%%%%%%%%%%%%%%%%%%%%%%%%%%%%%%%%%%%
%% MAINMATTER
%%%%%%%%%%%%%%%%%%%%%%%%%%%%%%%%%%%%%%%%%%%%

\section{Introduction}

One of the fundamental goals of the modern Cosmology is to measure the distance of the emitting systems, a key parameter
linked to the size, energy and mass of them. In particular, extragalactic distance measurements are fundamental in the 
learning of the size, age, composition and evolution of the Universe \cite{REvDIst1, REvDIst2}.
Depending on the distance of the object, the method of the measure changes radically. There are several problems in the 
cosmic distance ladder due to this approach, for which we can not discriminate the best parametrization (in term of dark 
energy and dark matter) \cite{REvDIst2}. 
Among these, the main one is that the errors and the uncertainties increase at each steps. Another one is linked to the 
calibration: in fact, we have to determinate a class of object with absolute magnitude very well-known. We could resolve a 
part of these problems using another type of distance indicator for which
we don not use the electromagnetic flux but the gravitational one. Here we show that the coalescing binary systems could 
be used to determine independently the distances, so we can reduce the errors and the uncertainties and we can improve 
the calibration.

%%%%%%%%%%%%%%%%%%%%%%%%%%%%%
\section{Coalescing binaries system as standard candles}
%%%%%%%%%%%%%%%%%%%%%%%%%%

A binary system is composed of two stars (neutron stars (NS), white
dwarves (WD) and black holes (BH)) orbiting around their
center of mass, whose components inspiral and
loss angular momentum and energy via gravitational waves (GWs) emission.
As a consequence of the coalescence GWs frequency increases and, if it is observed, 
it could be a "signature" for these dynamic systems. For this reason they are usually considered strong
emitter of GWs and they could be extremely useful for cosmological distance ladder if the physical features of GWs
emission are well determined.
For the coalescing systems, we can define:
\begin{equation}
M_c =\frac{(m_1m_2)^{3/5}}{(m_1+m_2)^{1/5}},
\end{equation}
a particular combination of the masses of the two stars, known as the \emph{chirp mass}. By determining the
amplitudes for the two polarizations of the GWs at lowest order in $v/c$ \cite{Maggiore},
and by averaging over the orbital period and the orientation of the binary orbital plane,
we can write the average (characteristic) GWs amplitude:
\begin{equation}\label{eq:Dis}
h_c \left( t \right) = \frac{4}{{r \left( z \right)}}\left( {\frac{{G {\cal{M}}_c \left( z \right)}}{{c^2 }}} \right)^{5/3} \left( {\frac{{\pi f \left( t \right)}}{c}} \right)^{2/3}\, .
\end{equation}
Eq. \eqref{eq:Dis} has to be modified if we consider a binary at a cosmological distance,
i.e. at redshift $z$. By considering the propagation of the GWs in a Friedmann-Robertson-Walker Universe
and by taking into account that all quantities appearing in Eq. \eqref{eq:Dis} are measured
by the observer, we have simply to apply the redshift correction (due to expansion of the Universe), as follows:
\begin{enumerate}
\item The frequency $f(t)$ is replaced by $f_{\rm obs}$, which is the frequency measured by the observer.
 It is redshifted with respect to the source frequency $f_s$, i.e. $f_{\rm obs}=f_s/(1+z)$;
\item the chirp mass $M_c$ must be replaced by ${\cal M}_c =(1+z) M_c$;
\item the distance $r$ to the source must be replaced by the luminosity distance $d_L(z)$.
\end{enumerate}
If we are able to measure the quantities related to GWs emission ($h_c(t)$,$f(t)$) and if we can estimate 
the chirp mass and the redshift ($z$) then we can determine the cosmological distance $d_L$, founding a 
gravitational standard candle and testing the cosmological model \cite{Schutz0}.
For the determination of the redshift $z$, several possibilities have been proposed, like for example the 
searching of an optical counterpart. In particular, short GRBs appear related to such systems and they are 
quite promising potential GWs standard sirens \cite{Dalal, noi}. Another possibility, adopted in this work, 
is that the redshift of the binary systems can be associated to the barycenter of the galaxy or the galaxy 
cluster hosting the systems.
%%%%%%%%%%%%%%%%%%%%%%%%%%%%%%%%%%%%%%%%%%
\section{Numerical analysis and results}
%%%%%%%%%%%%%%%%%%%%%%%%%%%%%%%%%%%%%%%%%%

We consider coalescing binary systems at redshifts (without uncertainties) $z < 0.1$ without any systematic 
errors in order to obviate the absence of a complete catalogue of such systems. The choice of low redshifts 
is due to the observational limits of ground-based interferometers like VIRGO or LIGO \cite{VIRGO, LIGO}. 
In the simulation here presented, the sources are slightly out of LIGO-VIRGO observable range, but in 
principle, they are within future improvements like Advanced VIRGO and Advanced LIGO.

Here, we have used the redshifts taken by NASA/IPAC  EXTRAGALACTIC  DATABASE (NED) \cite{Abell},
fixing them at the barycenter of the host galaxy/cluster. The binary
\emph{chirp mass} $M_C$ is  typically measured from the newtonian
part of the signal at upward frequency sweep, to $\sim 0.04\%$ for
a NS/NS binary and $\sim 0.3\%$ for a system containing at least
one BH \cite{original, Cutler}. The luminosity distance of the binary system
can be inferred, from the observed waveforms, to a precision $\sim 3/ \rho \leq
30\%$, where $\rho = S/N$ is the amplitude signal-to-noise ratio
in the total LIGO network. By fixing the characteristic amplitude of GWs,
we tune the frequencies in a range compatible  with such a fixed
amplitude, then the error on distance luminosity is calculated by the error
on the chirp mass with standard errors propagation.

\begin{figure}
 \begin{minipage}[b]{7.5cm}
   \centering
   \includegraphics[width=0.98\columnwidth]{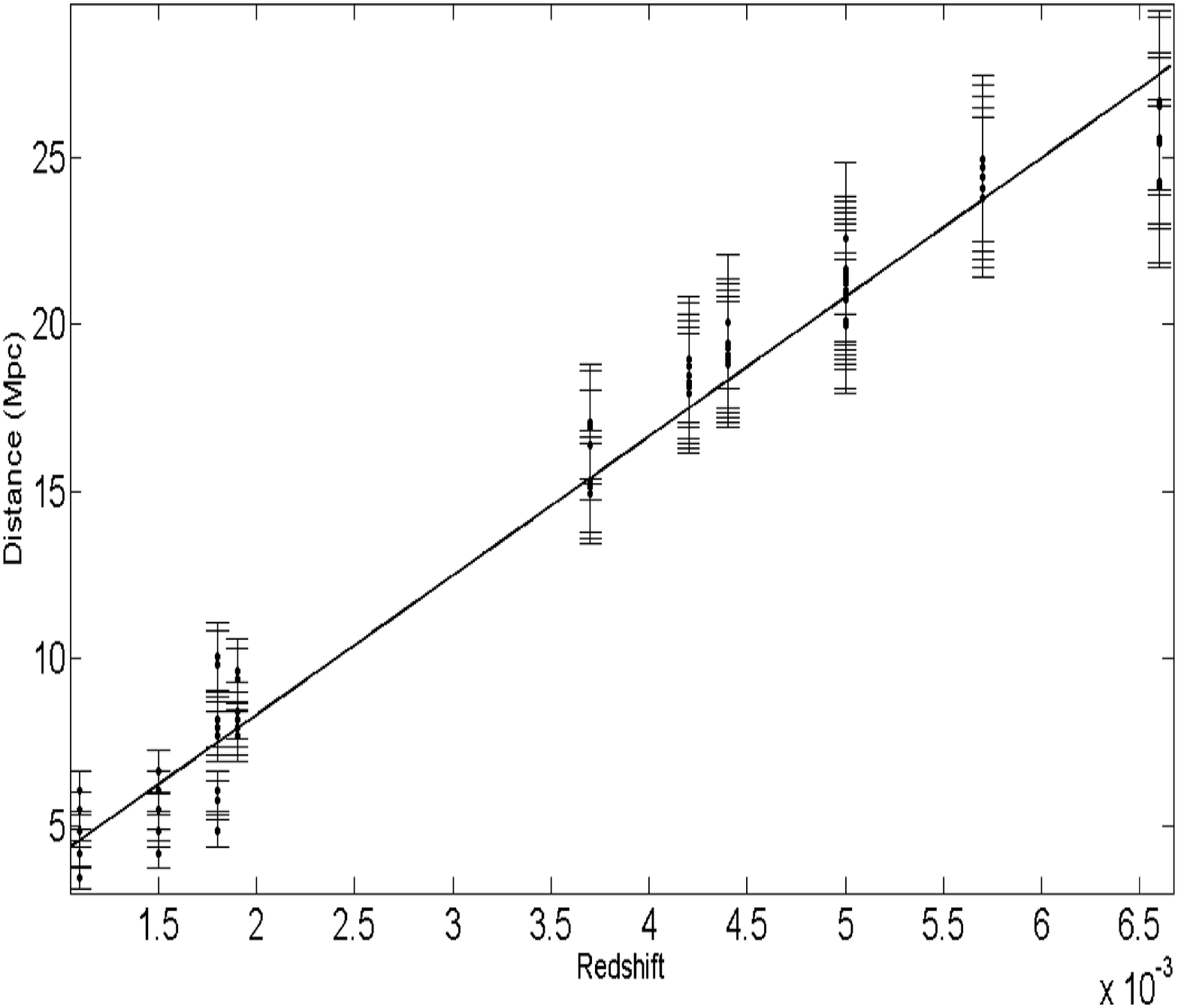}
   \caption*{\textbf{(a)} NS-NS systems.}
 \end{minipage}
 \ \hspace{3mm} \
 \begin{minipage}[b]{7.5cm}
  \centering
   \includegraphics[width=0.98\columnwidth]{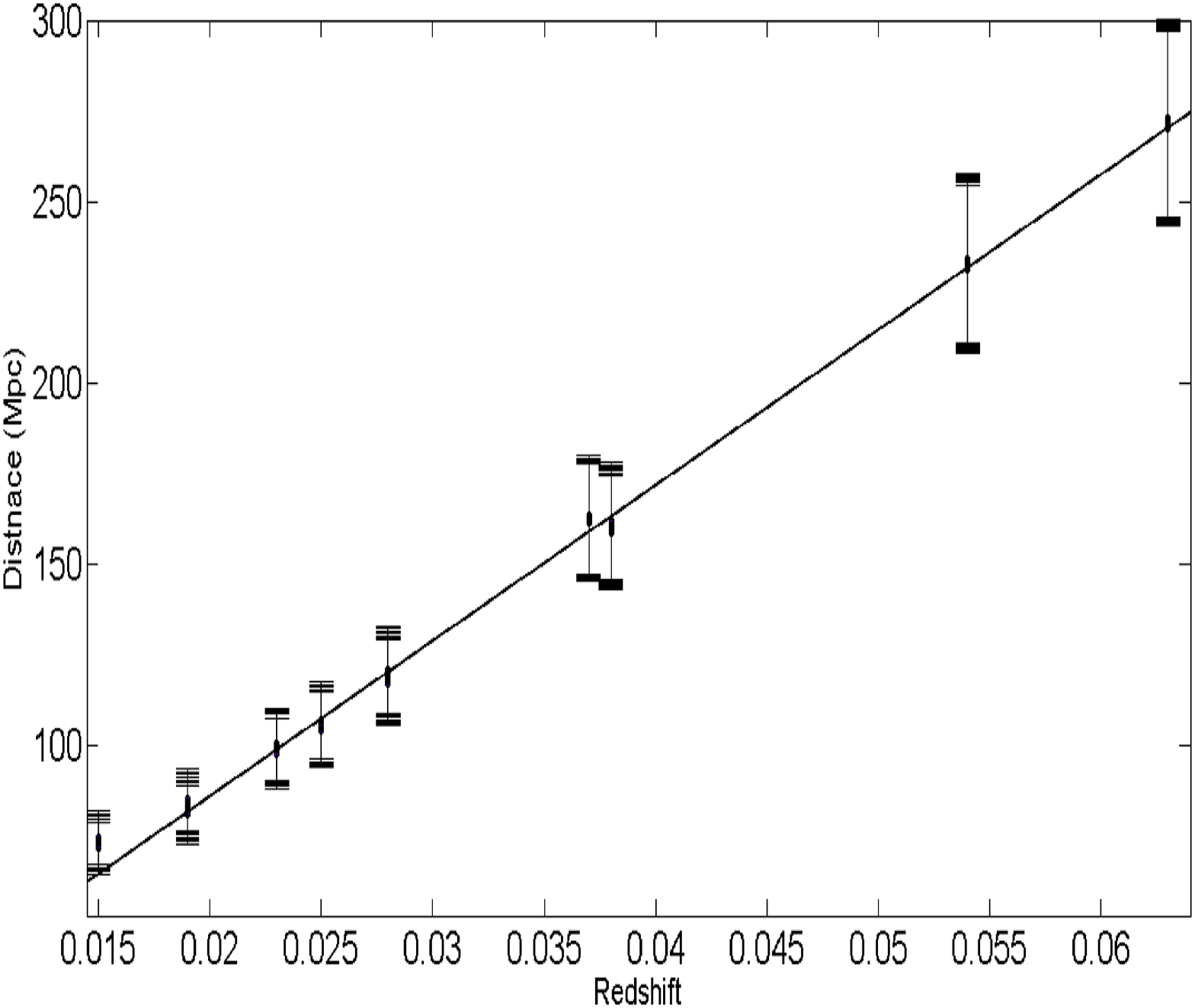}
   \caption*{\textbf{(b)} BH-BH systems.}
 \end{minipage}
\caption{Luminosity distance vs redshift for simulated systems.}
\label{fig:plot}
\end{figure}

The systems here considered are NS-NS and BH-BH. For each of them, a particular
frequency range and a characteristic amplitude (beside the chirp
mass) are fixed. We start with the analysis of NS-NS systems ($M_C
= 1.22 M_{\odot}$) with characteristic amplitude fixed to the
value $10^{-22}$. In Table \ref{tab:tabella}(a), we report the
redshift, the value of $h_C$ and the frequency range of systems
analyzed. In Fig. \ref{fig:plot}(a), the derived Hubble relation is
reported. The Hubble constant value  is $72\pm1$ km/s Mpc in agreement with
the recent WMAP  estimation \cite{WMAP7}. The same procedure is adopted for BH-BH systems
($M_C=8.67 M_{\odot}$, $h_C=10^{-21}$). The data and the results are respectively in
Table \ref{tab:tabella}(b) and in Fig. \ref{fig:plot}(b), and the Hubble constant
value computed by these systems is $69\pm2$ km/s Mpc. In this case the error is
larger than the previous one, because we have a larger indetermination
in the measure of the mass of the BH.

\begin{table}
 \begin{minipage}[b]{6.5cm}
\centering
  \begin{tabular}{|c|c|c|c|}
  \hline
  \textbf{Object} & \textbf{z} &
  \textbf{$h_c$} & \textbf{Freq.} \\
  & & &  (Hz)\\[1ex]
  \hline
  NGC 5128 & 0.0011 & $10^{-22}$ & $0\div10$\\[1ex]
  NGC 1023 Gr.  & 0.0015 & $10^{-22}$ & $0\div10$\\[1ex]
  NGC 2997 & 0.0018 & $10^{-22}$ & $5\div15$\\[1ex]
  NGC 5457 & 0.0019 & $10^{-22}$ & $10\div20$\\[1ex]
  NGC 5033 & 0.0037 & $10^{-22}$ & $25\div35$\\[1ex]
  Virgo  Cl.& 0.0042 & $10^{-22}$ & $30\div40$\\[1ex]
  Fornax Cl. & 0.0044 & $10^{-22}$ & $35\div45$\\[1ex]
  NGC 7582 & 0.0050 & $10^{-22}$& $ 45\div55$\\[1ex]
  Ursa Major Gr. & 0.0057 & $10^{-22}$& $50\div60$\\[1ex]
  Eridanus Cl. & 0.0066 & $10^{-22}$ & $55\div65$\\
  \hline
    \end{tabular}
\caption*{\textbf{(a)} NS-NS systems.}
 \end{minipage}
\ \hspace{3mm} \
\begin{minipage}[b]{6.5cm}
\centering
  \begin{tabular}{|c|c|c|c|}
  \hline
  \textbf{Object} & \textbf{z} &
  \textbf{$h_c$} & \textbf{Freq.} \\
  & & &  (Hz)\\[1ex]
  \hline
  NGC 5128 & 0.0011 & $10^{-22}$ & $0\div10$\\[1ex]
  NGC 1023 Gr.  & 0.0015 & $10^{-22}$ & $0\div10$\\[1ex]
  NGC 2997 & 0.0018 & $10^{-22}$ & $5\div15$\\[1ex]
  NGC 5457 & 0.0019 & $10^{-22}$ & $10\div20$\\[1ex]
  NGC 5033 & 0.0037 & $10^{-22}$ & $25\div35$\\[1ex]
  Virgo  Cl.& 0.0042 & $10^{-22}$ & $30\div40$\\[1ex]
  Fornax Cl. & 0.0044 & $10^{-22}$ & $35\div45$\\[1ex]
  NGC 7582 & 0.0050 & $10^{-22}$& $ 45\div55$\\[1ex]
  Ursa Major Gr. & 0.0057 & $10^{-22}$& $50\div60$\\[1ex]
  Eridanus Cl. & 0.0066 & $10^{-22}$ & $55\div65$\\
  \hline
    \end{tabular}
\caption*{\textbf{(b)} BH-BH systems.}
 \end{minipage}
\caption {For each clusters we indicate redshifts, characteristic amplitudes, frequency range for coalescing binary systems.}
\label{tab:tabella}
\end{table}

\section{Conclusions}

We have considered simulated binary systems,
whose redshifts can be estimated considering the barycenter of the
host astrophysical system as  galaxy,  group of galaxies or
cluster of galaxies.
We have simulated
NS-NS and BH-BH binary systems. Clearly, the leading
parameter is the chirp mass $M_c$, or its red-shifted counter-part
${\cal M}_c$, which is directly related to the GWs amplitude. The
adopted redshifts are in a  well-tested  range of scales and  the
Hubble constant value is in good agreement with WMAP estimation.
The Hubble-luminosity-distance diagrams of the above simulations
show the possibility to use the coalescing binary systems as
distance indicators. The limits of the method are, essentially,
the measure of GWs polarizations and the redshifts but it could lead to completely
independent determinations and it could complement and increase the
confidence of other standard candles \cite{HH}.
\bibliographystyle{plain}

\end{document}